\documentstyle[times,pramana,epsfig,floats]{ias}
\def\beq{\begin{equation}}
\def\eeq{\end{equation}}
\def\bea{\begin{eqnarray}}
\def\eea{\end{eqnarray}}

\newcommand{\cC}{{\cal C}}

\def \np{{ Nucl.\ Phys.~\/}}
\def \pl{{ Phys.\ Lett.~\/}}
\def \pr{{ Phys.\ Rev.~\/}}

\newcommand{\diracslash}[1]{#1\llap{/\kern2pt}}

\newcommand{\beqy}{\begin{eqnarray}}
\newcommand{\eeqy}{\end{eqnarray}}
\newcommand{\ba}[1]{\begin{array}{#1}}
\newcommand{\ea}{\end{array}}

\begin{document}
\mark{{Heavy-ion physics and QGP }{M. G. Mustafa and S. Raniwala}}
\title{ Working Group Report: Heavy Ion Physics and Quark-Gluon Plasma}

\author{Coordinators: MUNSHI G. MUSTAFA$^1$ and SUDHIR RANIWALA$^2$}
\author{{\it Contributors}: T. Awes$^3$, B. Rai$^4$, R. S. Bhalerao$^5$, 
J. G. Contreras$^6$, R. V. Gavai$^5$, S. K. Ghosh$^7$, P. Jaikumar$^8$, 
G. C. Mishra$^9$, A. P. Mishra$^4$, H. Mishra$^{10}$, B. Mohanty$^{11}$, 
J. Nayak$^{11}$, J-Y. Ollitrault$^{12}$, S. C. Phatak$^4$, 
L. Ramello$^{13}$, R. Ray$^1$, A. K. Rath$^4$, 
P. K. Sahu$^4$, A. M. Srivastava$^4$, D. K. Srivastava$^{11,14}$, 
V. K. Tiwari$^{15}$}
\address{$^1$Theory Group, Saha Institute of Nuclear Physics, 
1/AF Bidhan Nagar, Kolkata 700 064, India\\
$^2$University of Rajasthan, Jaipur 302004, India\\ 
$^3$ Oak Ridge National Laboratory, High Energy Physics,
Oak Ridge, Tennessee 37831-6372, USA \\
$^4$Institute of Physics, Sachivalaya Marg, Bhubaneswar 700 005,
India\\
$^5$Tata Institute of Fundamental Research, Homi Bhabha Road,
Mumbai 400 005, India\\ 
$^6$ Departamento de Fisica Aplicada, CINVESTAV, M\'erida, 
Yucat\'an, M\'exico and PH Division, Centre for European Nuclear Research, Geneva, Switzerland\\
$^7$Department of Physics, Bose Institute, 93/1, A. P. C. Road,
Kolkata 700 009, India\\
$^8$Physics Division, Argonne National Laboratory, Argonne, 
IL 60439-4843, U.S.A.\\
$^9$ Sikkim Manipal Institute of Technology, Majitar, Rango-737132,
East Sikkim, Sikkim\\
$^{10}$Theory Division, Physical Research Laboratory,
Navrangpura, Ahmedabad 380 009, India\\
$^{11}$Variable Energy Cyclotron Centre, 1/AF Bidhan Nagar, Kolkata
700 064, India\\
$^{12}$ Service de Physique Th\'eorique, CEA/DSM/SPhT, Unit\'e de 
Recherche associ\'{e}e au CNRS, F-91191 Gif-sur-Yvette Cedex, France\\
$^{13}$ Universit\`a del Piemonte Orientale, Alessandria and 
INFN-Torino, Italy\\ 
$^{14}$ Van de Graff Laboratory, Nuclear Physics Division, Bhabha Atomic 
Research Centre Trombay, Bombay - 400 085, India.\\
$^{15}$ Physics Department, Allahabad University, Allahabad 211002, 
India\\
\sf{E-Mail:}munshigolam.mustafa@saha.ac.in , raniwala@uorehep.ac.in}

\abstract{This is the report of Heavy Ion Physics and Quark-Gluon Plasma
at WHEPP-09 which was part of Working Group-4.  Discussion and work on
some aspects of Quark-Gluon Plasma believed to have created in heavy-ion 
collisions and in early universe are reported.}

\maketitle

\section*{Introduction}
\label{intro}

Soon after the discovery of Quantum Chromodynamics (QCD), it was 
conjectured 
that at high temperature $T$ the color charge is screened~\cite{Shuryak78}
and the corresponding phase of matter was named Quark-Gluon Plasma (QGP).
It is a special kind of plasma in which the electric charges are replaced by
the color charges of quarks and gluons, mediating the strong interaction
among them. Such a state of matter is expected to exist at extreme 
temperatures, above $150$ MeV, or densities, above about $10$ times normal 
nuclear matter density. These conditions could existed in the early 
universe for the first few microseconds after the Big Bang or in the interior
of neutron stars. 

The aim of the ongoing relativistic heavy-ion collision experiments is
to explore the possible QGP phase of QCD. The essential difference 
in heavy-ion collisions over the nucleon-nucleon collisions is the 
dominance of the partonic-level description for essentially all momentum 
scales and over nuclear size distances. In order to describe the produced
system as a state of matter it is necessary to establish that these  
nonhadronic degrees of freedom form a statistical ensemble. Therefore,
the concepts such as temperature, chemical potential and flow velocity 
should apply and the system can be explained by an experimentally determined 
equation of state. In addition experiments should eventually be able to 
determine the physical characteristics of the phase transition, {\it viz.}, 
the critical temperature, the order of the phase transition, and the speed
of sound along with the nature of underlying quasi-particles. To date such
description is provided by the Lattice QCD (LQCD) calculations. 
Ultimately, one would expect to validate this by characterizing the QGP in 
terms of its experimentally observed properties.
The commissioning of Relativistic Heavy Ion Collider (RHIC) at BNL
and various experiments performed
therein have ushered in a new era. These experiments have acquired data on
{\it Au+Au} collisions at various energies, an essential {\it p+p} baseline
data set, and a critical {\it d+Au} comparison. The analyses of these various
system have yielded a rich abundance of results.

There were non-overlapping talks and overlapping talks during WHEPP. The non-
overlapping talks on this subject 
were focussed on the 
recent developments in LQCD, present status of experimental 
observations and the corresponding theoretical efforts, 
and certain aspects of future experiments, 
 in understanding 
the properties of QGP produced in heavy-ion collisions. 
Details are reported in these proceedings.
 
\begin{enumerate}
\item Status of Lattice QCD: {\it Rajiv V. Gavai}.
\item Results from STAR Experiment at RHIC: {\it Bedanga Mohanty}.
\item PHENIX Overviews - Status of QGP: {\it Terry Awes}.
\item Jets in ALICE: {\it J. Guillermo Contreras}.
\item Theoretical Status of QGP: {\it Jean-Y. Ollitrault}.
\item High Energy Photons from Relativistic Heavy Ion Collider: {\it Dinesh K. 
Srivastava}.
\item On the Structure and Appearance of Quarks Stars: {\it Prashanth 
Jaikumar}.
\end{enumerate}

Recent numerical LQCD calculations has given us wealth of information 
on various thermodynamic properties at finite temperature and chemical
potential. It has now been established that there
is only a {\it crossover}~\cite{edwin} of normal hadronic matter to a state 
of deconfined
quarks and gluons at temperature $T_c\sim 200$ MeV. Moreover, {\it the equation 
of state}~\cite{edwin}, {\it various susceptibilities}~\cite{alton,sourendu} and
{\it transport coefficients}~\cite{transport} have been obtained. 
It is also found  that 
charmonium states remain bound at least up to $T\sim 2T_C$~\cite{Datta} 
and the behavior of temporal correlators in pseudoscalar and vector 
channels deviates 
significantly from the free behavior at $T\sim 3T_C$~\cite{Karsch}. These 
analyses suggested {\it that pseudoscalars and vectors resonances may exist 
above $T_C$}. 

Robust results from  {\it Au+Au} at the BNL RHIC 
experiments have shown collective
effects known as radial~\cite{radial} and elliptic~\cite{elliptic,elliptic1}
flows, and a suppression of high-$p_{T}$ hadron spectra~\cite{elliptic1,RHIC},
which could possibly indicate the quenching of light quark and gluon
jets~\cite{Pluemer}. The hydrodynamical description of the
observed collective flow indicates that the matter produced at RHIC {\it
behaves like a near-perfect fluid}~\cite{fluid}.
On the other hand the amount of jet quenching might depend 
on the state of matter of the fireball, i.e., QGP or a hot hadron gas. There 
are extensive theoretical efforts in understanding the 
effect of the medium on jet quenching~\cite{Gyulassy,Salgado,Zakharov,Mueller,Munshi} using
collisional as well as radiative energy loss since the high energy partons
traveling through a medium will lose energy owing to the interactions
in the medium. 
The measurement in RHIC~\cite{photon} indicates 
{\it excess direct photons} over the next 
to
leading order pQCD processes in RHIC has also been 
reported and  extensive theoretical efforts were made to 
understand these excess photons~\cite{srivastava1}. 
In astrophysical part
the focus was to study and understand
the {\it structure} of Standard Model at low energy and  the {\it evolution}
 of the early Universe.

The working group activity was structured around review talks followed
by discussions. The emphasis in this working group (IV) was to establish
connection between the subject matter of the review talks in plenary sessions
and some informal talks during discussion sessions, which led to identification
of relevant physics problems along with efforts to work them out. This working 
group IV was particularly interested in following topics and relevant problems:
{\it 1) Charmonium suppression,
2) Jet quenching and jet identification,
3) Hydrodynamic description of elliptic flow,
4) Susceptibilities and speed of sound in QGP,
5) Possibility of bound states in QGP,
6) $\phi$-production at RHIC,
7) AdS/CFT -- QCD/QGP  and
8) Neutrino emission from crystalline color superconducting quark matter.}

In the next few sections we will briefly report the discussion held on 
these
topics, and also the progress made on the relevant physics problems 
undertaken by the working group members.

\section{Charmonium suppression}
\label{jpsi}
{\it L. Ramello}

\vspace{0.2in}
The latest results from CERN SPS fixed target experiments NA50 and NA60 about 
heavy quarkonia (mostly charmonium) production in p-nucleus collisions at 400 
and 450 GeV and in In-In, Pb-Pb collisions at 158 GeV/nucleon are presented.
For a historical overview of charmonium studies at SPS see \cite{kluberg}

Experiments NA50 and NA60 share the same muon spectrometer which allows to 
trigger on dimuons emitted from charmonium and bottomonium decays, as well
from the Drell-Yan process and open charm associated production.
NA50 has three independent centrality detectors: an electromagnetic calorimeter, a
multiplicity detector and a zero-degree calorimeter (ZDC). NA60 retains the 
ZDC for centrality measurement and has a vertex magnet and a silicon pixel 
vertex spectrometer which allows matching between the muon tracks in the muon
spectrometer and in the vertex spectrometer.

The recent NA50 p-A results at 450 and 400 GeV \cite{na50-pA} allow (together with 
NA38 and NA3 data at 200 GeV) to determine the J/$\psi$ nuclear absorption only 
from proton induced reactions, without reference to S-U results. A value of 
$\sigma_{absorption}$(J/$\psi$) = 4.18 $\pm$ 0.35 mb is obtained. This in turn allows to 
calculate the expected J/$\psi$ yield in Pb-Pb collisions at 158 GeV/nucleon and 
in S-U collisions at 200 GeV/nucleon.
The centrality dependence in Pb-Pb is calculated with a Glauber model using 
the variable L (average path in nuclear matter) as a common parameter in p-A,
S-U, In-In and Pb-Pb collisions.

The NA50 (J/$\psi$ / Drell-Yan) and ($\psi'$ / Drell-Yan) cross-section ratios as a 
function of centrality are obtained from Pb-Pb data samples collected at 158 
GeV per nucleon in the 1998 run and, under improved experimental conditions, 
in 2000 \cite{na50-pbpb}.

The J/$\psi$ expected production extrapolated from p-A data is then compared to 
the NA50 results in Pb-Pb collisions, as well as to the NA38 results for S-U 
reactions.
A departure from normal nuclear absorption is observed for mid-central Pb-Pb
collisions, with the suppression increasing with centrality. The three 
centrality variables give a consistent picture. On the other hand, peripheral
Pb-Pb data and all S-U data are compatible with normal nuclear absorption.
The $\psi'$ has a significant absorption in p-A collision, which
is further increased by about a factor 3 both in S-U and Pb-Pb collisions.
The J/$\psi$ suppression has been studied as a function of p$_T$: the anomalous
suppression is concentrated at low p$_T$. For p$_T > $ 3.5 GeV/c, the central to 
peripheral ratio R$_{CP}$ becomes almost independent of the centrality range. 
The average square p$_T$ (or equivalently the effective temperature) of the 
J/$\psi$ first increases with centrality and then saturates in central Pb-Pb 
collisions.

\begin{figure}[htbp]
\epsfxsize=12cm
\centerline{\epsfbox{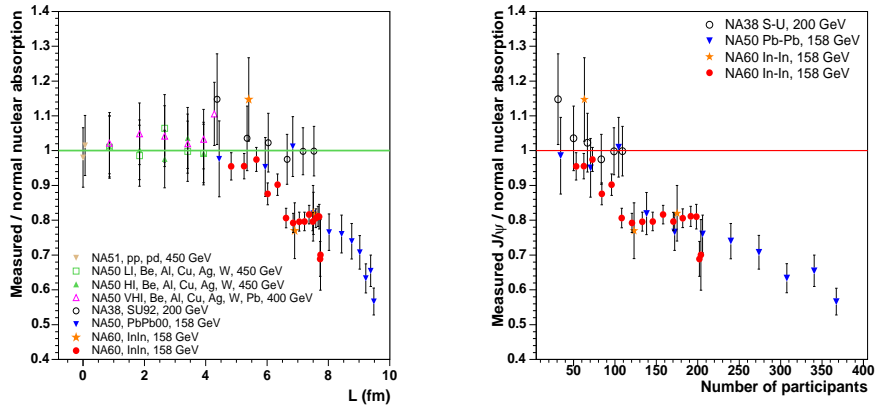}}
\caption{The S-U, In-In and Pb-Pb data points do not overlap in the L variable, as seen in the left panel. The J/$\psi$ suppression patterns for different interacting systems are in fair agreement with the N$_{part}$ variable.}
\label{ramello}
\end{figure}

The NA60 experiment has studied the centrality dependence of the J/$\psi$
production in another system, namely In-In, at 158 GeV per nucleon \cite{na50-inin}.
The J/$\psi$/Drell-Yan ratio in In-In in three centrality bins compares 
well with NA50 Pb-Pb results, in the sense that anomalous suppression is
seen also in In-In in the limited range of L accessible to this analysis.
NA60 has presented more detailed (still preliminary) results by directly 
comparing the measured J/$\psi$ sample with a theoretical distribution expected in
the case of pure nuclear absorption. 
The onset of anomalous suppression in In-In is clearly seen in the range
80 $<$ N$_{part}$ $<$ 110 (N$_{part}$ being the number of participating nucleons), with 
saturation for larger values of N$_{part}$, up to about 200.
Comparing the NA60 direct J/$\psi$ analysis with NA38/NA50 results, it is 
evident that the three data sets do not overlap when plotted against the 
L variable, while a rather good overlap is seen when they are plotted
against the N$_{part}$ variable, as seen in Fig.~\ref{ramello}.

The SPS data presented above and the PHENIX results \cite{tram} obtained with Au+Au
and Cu+Cu collisions at RHIC (c.m. energy of 200 GeV/nucleon) have been
compared with a few models \cite{satz,rapp,capella,kostyuk}, which attempt to describe data 
under different assumptions.
Models have in general been tuned on p-A, S-U and Pb-Pb SPS data and give 
predictions for In-In data and for RHIC Au-Au (Cu-Cu) data.

Model \cite{satz}, based on $\chi_c$ suppression in a percolation scenario, predicts an 
onset of the suppression at SPS in Pb-Pb collisions at N$_{part} \sim$ 125, which 
agrees with NA50 data, and an onset in In-In collisions at N$_{part} \sim$ 140, which
disagrees with NA60 data (the onset in data is at N$_{part} \sim$ 90).

Model \cite{rapp} contemplates charmonium suppression and regeneration in both QGP
and Hadron Gas phases. At SPS only a small amount of regeneration is needed,
and both NA50 and NA60 data are well described, except for In-In data at 
N$_{part} >$ 160 which are underpredicted. At RHIC 200 GeV c.m. energy in Au-Au
collision a substantial amount of regeneration is needed to describe data.

Model \cite{capella} includes nuclear absorption ($\sigma_{abs}$ = 4.5 mb) and comover absorption
($\sigma_{co}$ = 0.65 mb). SPS data for Pb-Pb collisions are well described, while 
those for In-In collision are slightly underpredicted: the ratio Measured/Expected
(expected from normal nuclear absorption) for semi-central and central collisions
is 0.80 in data with no centrality dependence, 0.70 and decreasing in the model.
The R$_{AA}$ ratio at RHIC for Au-Au and Cu-Cu collisions at N$_{part} \sim$ 300 is largely
underpredicted (0.10 vs. 0.35).

Model \cite{kostyuk} has in fact two scenarios: the QGP scenario underpredicts the Au-Au PHENIX
data, while the Statistical Coalescence Model (valid for N$_{part} >$ 100) predicts a
flat behaviour vs. N$_{part}$ and roughly agrees with PHENIX dN(J/$\psi$)/dy.

\section{Jets}
\label{jet}
\subsection{Jets in ALICE}
\label{jetsi_alice}
{\it J.G.Contreras}

\vspace{0.2in}
Jets, at the parton level, are cascades of quarks and gluons produced by a
fast moving parton in a process called fragmentation. In the ALICE experiment,
the fast partons are created as a product of a hard scattering, where the
incoming partons are provided by the collision of two extremely
relativistic
protons or nuclei. At the detector level, jets are the highly collimated
bundle of particles produced by the hadronization process of the partonic
jet.

The fragmentation properties of a parton jet are influenced by its
interaction
with the medium, where the parton loses its energy in the QGP. 
This phenomenom is called jet quenching and it has been
advocated
to be an ideal probe of the QGP state created by the collision of two
heavy
ions at asymptotic energies \cite{Pluemer}.
Partonic jets from a hard scattering are created before the QGP in
the collision of two heavy ions. Hard scatterings and the
evolution of fragmentation in vacuum can be
quite reliably computed within pQCD. Any changes in the
fragmentation process in the presence of QGP can be 
determined, in principle, from the measurement 
of jet quenching. There are some caveats in this approach.
The experiments measure jets after hadronization, the effect of which
can not be calculated a priori, inhibiting a unique determination of the 
effect of QGP on the partons. However, there are models which
assume (with certain confidence) that the influence of hadronization 
is small in certain regions of phase space. Also, lacking detailed 
understanding of the QGP state, the understanding of the interaction of
a parton jet with a QGP is still in an evolutionary phase.

While there have been attempts to understand the effect of the medium
on the jet quenching due to the radiative energy 
loss~\cite{Gyulassy,Salgado,Zakharov,Mueller}, 
and recently also due to the collisional energy loss~\cite{Munshi}. 
Furthermore, only the first approximation to radiative energy loss 
has been used, but a lot of theoretical work is under progress 
in this area. Not withstanding these limitations, observation of jet
quenching in RHIC data \cite{RHIC} has been used to 'announce' the 
creation of a QGP. 

The experiments at RHIC have shown that there is jet quenching and that it is
definitely a final state effect. It must be mentioned here that there is 
an inherent bias in leading particle analysis, which can be best eliminated
by studying the complete jets. Further, it is desirable 
to study the QCD evolution of jet quenching. This can be best done by studying 
evolution of jet quenching with the hardness of the scale by comparing jet
shapes in a wide kinematic domain. These steps lead to the ALICE experiment
at the Large Hadron Collider at CERN; to new problems, and also to 
new opportunities.

The physics reach of studying jets in the ALICE experiment has been
reported in \cite{Contreras} and a brief overview is discussed
here. 

It is expected that under nominal conditions of LHC running,
several tens of jets with energies below 15 GeV will be 
produced in each Pb-Pb event within the acceptance of ALICE.
It is also expected that there will be jet(s) of 250 GeV for one
month of data taking. The large charge particle 
multiplicity envisaged in the heavy ion 
collision, in contrast to pp collision, will lead to several hundreds of 
GeV of energy deposition in one unit in the azimuth-rapidity space. 
The fluctuations in the average energy in a given cone are of the 
order of several tens of GeV and form a formidable background to observe
jets.

The strategy planned for the ALICE experiment is to use smaller cones, 
and to search for the jets using  particles whose energy lies above a given threshold and to
iteratively subtract the background using a modified jet cone algorithm
\cite{Blyth1}. It has been shown that above 50-60 GeV of energy, jets and its
spectra
can be reliably reconstructed using this modified jet algorithm on the
charged particles produced in the interaction and measured by the TPC.

Jet quenching affects certain properties of the jet like amount of
radiation outside a given cone, jet
heating and the fragmentation function \cite{Salgado}.
The various variables have relative merit e.g. one variable (jet heating) does not 
evolve with the hardness of the interaction and can not be used to test the QCD
evolution of jet quenching, while remaining a good test of 
jet quenching at all hardness scales. In contrast, 
fragmentation function of jets offers a window to study the evolution
of jet quenching with the hardness of the interaction.

In summary, jet shapes are ideal testbeds to study jet quenching
phenomena. To use them it is mandatory to measure very low
energy particles and to understand with great precision the
contribution from the underlying event and its fluctuation.
ALICE fulfills all the requirements to perform these demanding
measurements and extract from them the properties of jet quenching
and thus increase our understanding of the QGP.

\subsection{Flow Coefficients and Jet Characteristics in Heavy Ion Collisions: New Methods of Jet Identification}
\label{jetalgo}
{\it S.C.Phatak}

\vspace{0.2in}
Identification of jets in heavy ion collisions is an important and challenging 
problem because modification of jet properties is expected to give information on
possible formation of quark-gluon plasma during the collision 
process~\cite{Gyulassy,Salgado,Zakharov,Mueller,Munshi}.  
Jet quenching has already been observed~\cite{RHIC}, although the observation 
is somewhat indirect. In particular, these studies are basically the correlation 
studies between the energetic hadrons which are expected to be the hadrons 
produced in a jet. Detecting and identifying jets in heavy ion collisions is an essential
prerequisite to study jet quenching. 

A new method for jet identification has been developed \cite{flow}. The method is 
based on the fact that 
the flow or Fourier coefficients for events containing jets have typical structure 
enabling identification of jet events and determining its opening angle 
and the number of particles. The even coefficients in back-to-back jets are observed
to be larger.

The flow coefficients are Fourier coefficients of the azimuthal distribution of particles 
produced in heavy ion collisions and are given by
\begin{equation}
v_{m,p_T}^2 = \int d\phi_1 d\phi_2 dp_{T1} dp_{T2} p_{T1} p_{T2} P(\phi_1) P(\phi_2) 
\cos m (\phi_1 - \phi_2)
\end{equation}
Given the experimental distribution of particles, the flow coefficients are given by $
v_{m,p_T}^2  =   \frac{1}{N^2} \sum_{i,j} p_{T,i} p_{T,j} \cos ( \phi_i - \phi_j)$
where $N$ is the number of particles in the event and $p_{T,i}$ is the transverse momentum
of $i^{th}$ particle.  
For uniformly distributed particles, all flow coefficients ( except for m=0 ) 
vanish where as
for $\delta$-function distribution all coefficients are unity. Thus, for an event with a jet
having a number of particles in a small azimuthal angle and the rest of the particles uniformly 
distributed, $v_m$'s are expected to be abnormally large. This can be used to identify a 
jet and to determine the jet properties. We have shown that\cite{paper}, if an event has jet
particles distributed in a narrow cone of azimuthal angle, we have
$v^2_{m,p_T}  =   \frac{N_j^2 <p_T>^2}{N^2} \Big [ 1 - m^2 \sigma^2 + {\cal O} ( m^4  ) 
\Big ]$
where $\sigma$ is the variance of the $\phi$-distribution and $<p_T>$ is the average 
transverse momentum carried by a hadron in the jet. Thus $N_j <p_T>$ is the transverse 
momentum of the jet. 

These expressions clearly suggest a method of obtaining jet properties from the flow 
coefficient. A linear fit to $v^2_{m,p_T}$, plotted as a function of $m^2$ 
would yield the number of jet particles and jet $p_T$ from the intercept on y axis and 
$\sigma$ from the slope. 

Let us now consider the case of two back-to-back jets in a background of uniformly
distributed particles. This case is of interest
because we expect that a hard parton scattering would produce such jets having equal and 
opposite jet momenta. We expect that quenching of one of the jets would broaden one of 
the jets and/or produce more jet particles. Thus, the characteristics of the two back-to-back 
jets would be different. Further, in an extreme situation, the fast moving parton of one of the
jets may be completely absorbed in the medium leading to removal of one of the jets. 
Assuming that the jet particles are distributed uniformly in the jet cone of the respective jets,
we get 
\begin{eqnarray}
v^2_{m, p_T} & = & \frac{1}{N^2} \Big [ <p_{T,1}> j_0(\frac{m \Delta \phi_{j1}}{2}) + (-1)^m 
<p_{T,2}> j_0(\frac{m \Delta \phi_{j2}}{2}) \Big ]^2
\end{eqnarray}
Here $N_{ji}$  are the jet particles in the $i$th jet, $\Delta \phi_i$ are the corresponding 
opening angles and $<p_{T,i}>$ the average transverse momenta of the respective jet particles. 
The presence of $(-1)^m$ factor in the second term above implies odd-even staggering in 
$v_m^2$'s. In particular, odd coefficients vanish for identical jets. This property would 
be useful in estimating the possible quenching of one of the jets. The method also works 
when transverse momentum is not measured and only the azimuthal angle of the particles is 
known.

The method was tested using simulations.  Jet particles were simulated by adding certain 
number of particles distributed uniformly in a jet cone of $\Delta \phi$ and having 
exponentially decreasing $p_T$ distribution and added to HIJING events\cite{hijing} 
adapted for LHC energies.
Different values of number of jet particles and 
jet opening angles were used for analysis. In the absence of any jet particles, the 
method yields $v^2_{m,p_T}$ values smaller than 0.05 Gev$^2$. 
In case of the data with jet only, the method outlined above is able to 
extract correctly the number of
jet particles, jet transverse momentum and jet opening angle.

\section {Dissipative relativistic hydrodynamics}
\label{hydro}
{\it R.S. Bhalerao}

\vspace{0.2in}
Certain global features of the RHIC data on ultrarelativistic
heavy-ion collisions have been successfully explained in the framework
of the boost-invariant ideal hydrodynamics: The main characteristics
of the observed elliptic flow ($v_2$) are reasonably well described by
ideal fluid dynamics, while requiring unreasonably large cross
sections in transport models. The ability of the former to reproduce
both the elliptic flow and single-particle spectra for measured
hadrons with $p_T\leq 2$~GeV/$c$ near midrapidity in minimum-bias
collisions is considered a significant finding at RHIC.
The dependence of the flow pattern on hadron masses
further supports the hydrodynamical picture.

However, the short time scale for equilibration is difficult to
account for microscopically \cite{baier}.  
(It has been argued recently that plasma
instabilities may provide a mechanism for fast thermalization.)
Moreover, several features of the data clearly signal the breakdown of
the ideal hydrodynamical description \cite{bhalerao}: 
Hydrodynamic models seem to work
for minimum-bias data but not for centrality-selected pion and
antiproton data. Indeed, the pion data presented in Fig. 36 of \cite{starres}
show that the measured $v_2$ in central collisions {\em overshoots\/}
the value obtained by a hydrodynamical computation, $v_2({\rm
data})>v_2({\rm hydro})$, contrary to the expectation that the $v_2$
measurements saturate the hydrodynamical limit. See also Fig. 6 of \cite{starres}
where the agreement between $v_2({\rm data})$ and $v_2({\rm hydro})$
is far from satisfactory. This indicates incomplete equilibration
and/or importance of dissipative effects.

Hydrodynamic calculations have shown that $v_2$ is independent of the
system size for a given shape of the collision zone. This is a
consequence of scale invariance of ideal hydrodynamics. However,
incomplete equilibration breaks this scale invariance making $v_2$
depend upon the system size. Certain dimensionless numbers like Mach
number, Knudsen number and Reynolds number can be used to characterize
the motion of the fluid, in particular its compressibility, degree of
thermalization and importance of viscous effects. Estimating these
numbers for Au-Au collisions at RHIC calls in question the assumption
of local equilibrium and applicability of ideal hydrodynamics, and at
the same time points to the need for dissipative relativistic
hydrodynamic description of nucleus-nucleus collisions at RHIC.

\section{Susceptibilities and speed of sound in QGP in PNJL model} 
\label{suscep}

{\it Sanjay K. Ghosh, Tamal K. Mukherjee, Munshi G. Mustafa and Rajarshi Ray}

\vspace{0.2in}

Susceptibility is the response of the system to an externally applied force. Quark 
number susceptibilities are the response of the quark number density
with the variation of chemical potential. 
Lattice data on QCD thermodynamics, particularly recent study of
higher order susceptibilities~\cite{alton,sourendu} have provided
valuable information and new insight about the properties of matter 
produced around and above the critical temperature, $T_C$. Perturbative
QCD calculations fail to describe the details of these results.
There are some model calculations to understand the physical picture of
the structures in higher order susceptibilities produced by those LQCD data.
The hadron gas resonance model~\cite{ejiri} describes the data well below $T_C$
but fails for $T>T_C$. The recently proposed~\cite{liao} scenario of colored
bound states also compares the data. However, their comparison is still 
not completely satisfactory. In this work\footnote{The
authors are thankful to Rajiv Gavai for useful discussion during WHEPP-09 where
this work was initiated. Full details of this work was reported in hep-ph/0603050
and published in Phys. Rev. D73, 114007 (2006).}
 we try to understand 
the features and structures of the susceptibilities produced by LQCD 
within the model described below.



The thermal average of the Polyakov-loop can be considered to be the 
order parameter for deconfinement transition \cite{polyl}. Hence a 
judicious use of the Polyakov-loop in effective models may prove to be
of great advantage. On the other hand, there are QCD inspired models, 
predominantly, the NJL model, has given rise to an interesting phase 
diagram in temperature and chemical potentials.

We study~\cite{ghosh} some of the thermodynamic properties of strongly 
interacting matter using the Polyakov-loop $+$ Nambu-Jona-Lasinio (PNJL)
model \cite{pnjl0}. The motivation behind the PNJL model is to couple
the chiral and deconfinement order parameters inside a single framework. 
We have computed the EOS, the quark number susceptibilities, 
the specific heat $C_V$, the speed of sound (basically its square, $v_s^2$),
and the conformal measure $\cC=\Delta/\epsilon$, where 
$\Delta = \epsilon - 3P$ is the interaction measure and $\epsilon$ and 
$P$ are respectively the energy density and pressure of strongly 
interacting matter. Comparisons of these quantities with those obtained
on the lattice were made.

\begin{figure}
\begin{minipage}[h]{0.48\textwidth}
\centering{\includegraphics[height=0.7\textwidth,width=1.0\textwidth]{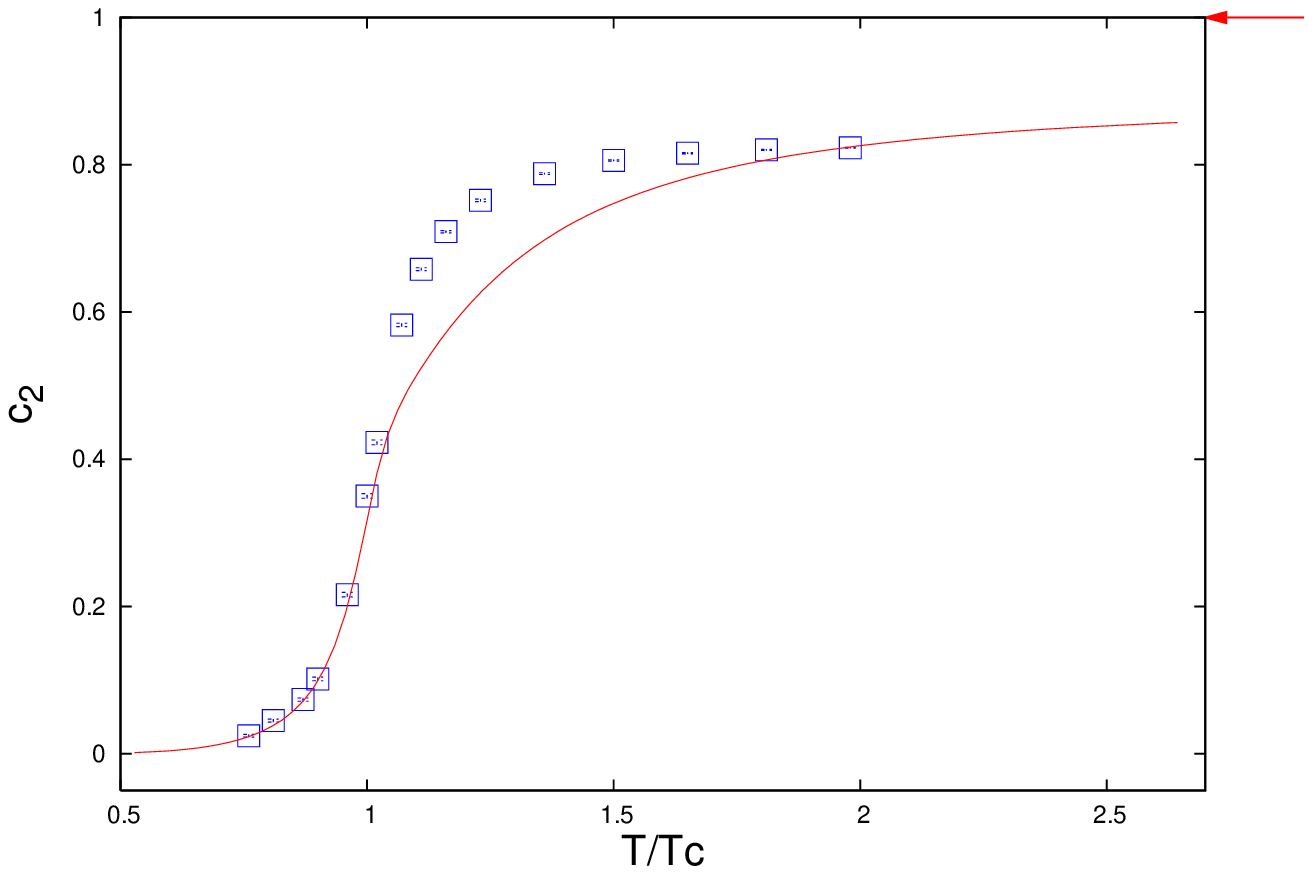}}
\end{minipage}
\begin{minipage}[h]{0.48\textwidth}
\centering{\includegraphics[height=0.7\textwidth,width=1.0\textwidth]{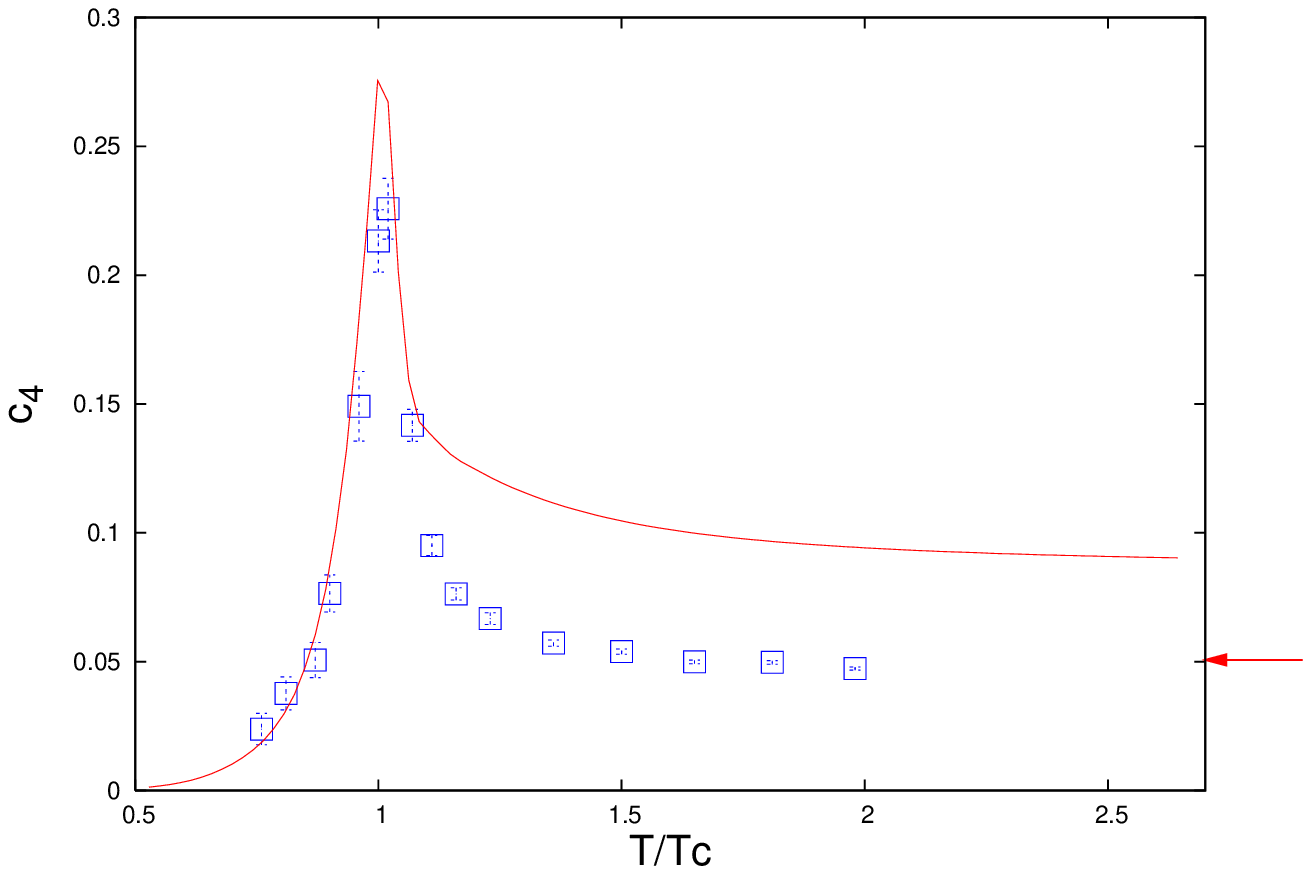}}
\end{minipage}
\caption{The QNS (left) and the forth order coefficient of reduced 
            pressure (right) as a function of $T/T_c$. Symbols are 
            Lattice data. Arrows on the right indicate 
            ideal gas values.}
\label{fg.scnord}\end{figure}


We use the parameterization of the PNJL model 
as given in Ref.\cite{pnjl2}. For our results we shall present the 
quantities as a function of temperature in units of a crossover 
temperature which is taken to be $T_c= \rm{227~MeV}$.

We found the pressure to grow from almost zero at low temperatures to 
about $90\%$ of its ideal gas value at $2.5 T_c$. Left panel in 
Fig.\ref{fg.scnord} shows the variation of the quark number susceptibility
(QNS), which is the second order coefficient ($c_2$) of reduced pressure 
($P/T^4$), with $T/T_c$. This shows an order parameterlike behavior. 
At higher 
temperatures the $c_2$ reaches almost 85 \% of its ideal gas value, 
consistent with Lattice data. Similar behavior has been observed in 
another model study~\cite{tamal} using density dependent quark mass 
model.

The fourth order derivative $c_4$, which can then be thought of as 
the "susceptibility" of $c_2$ shows a peak at $T=T_c$ (right panel of
Fig.\ref{fg.scnord}). Near the transition temperature $T_c$, the 
effective model should work well and we observe that the structure of 
$c_4$ is quite consistent with Lattice data. Just above $T_c$ however, 
there is a significant difference between our results of $c_4$ and that of 
Ref.\cite{alton}. Note that in the SB limit both $c_2$ 
and $c_4$ have only fermionic contributions. We expect that because the
coupling strength is still large in this temperature regime it is 
unlikely that $c_4$ should go to the SB limit within $T < 2.5 T_c$. 
Moreover, the quark masses used in Ref.\cite{alton} is considerably large
($m/T = 0.4$) to expect fermionic observables to go to the SB limit.
However, it is possible that our overestimation is due to the use of
mean field approximation. 

\begin{figure}
\begin{minipage}[h]{0.48\textwidth}
\centering{\includegraphics[height=0.7\textwidth,width=1.0\textwidth]{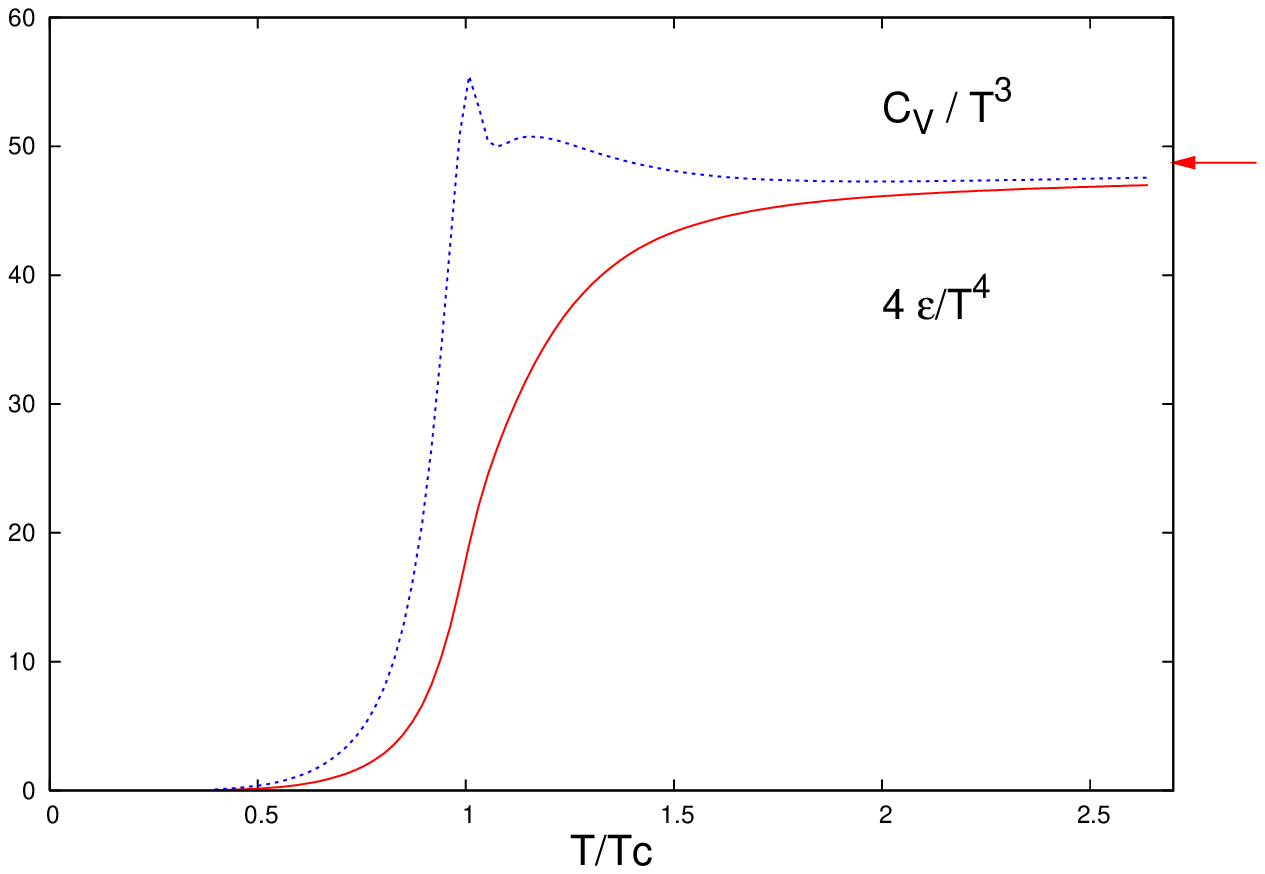}}
\end{minipage}
\begin{minipage}[h]{0.48\textwidth}
\centering{\includegraphics[height=0.7\textwidth,width=1.0\textwidth]{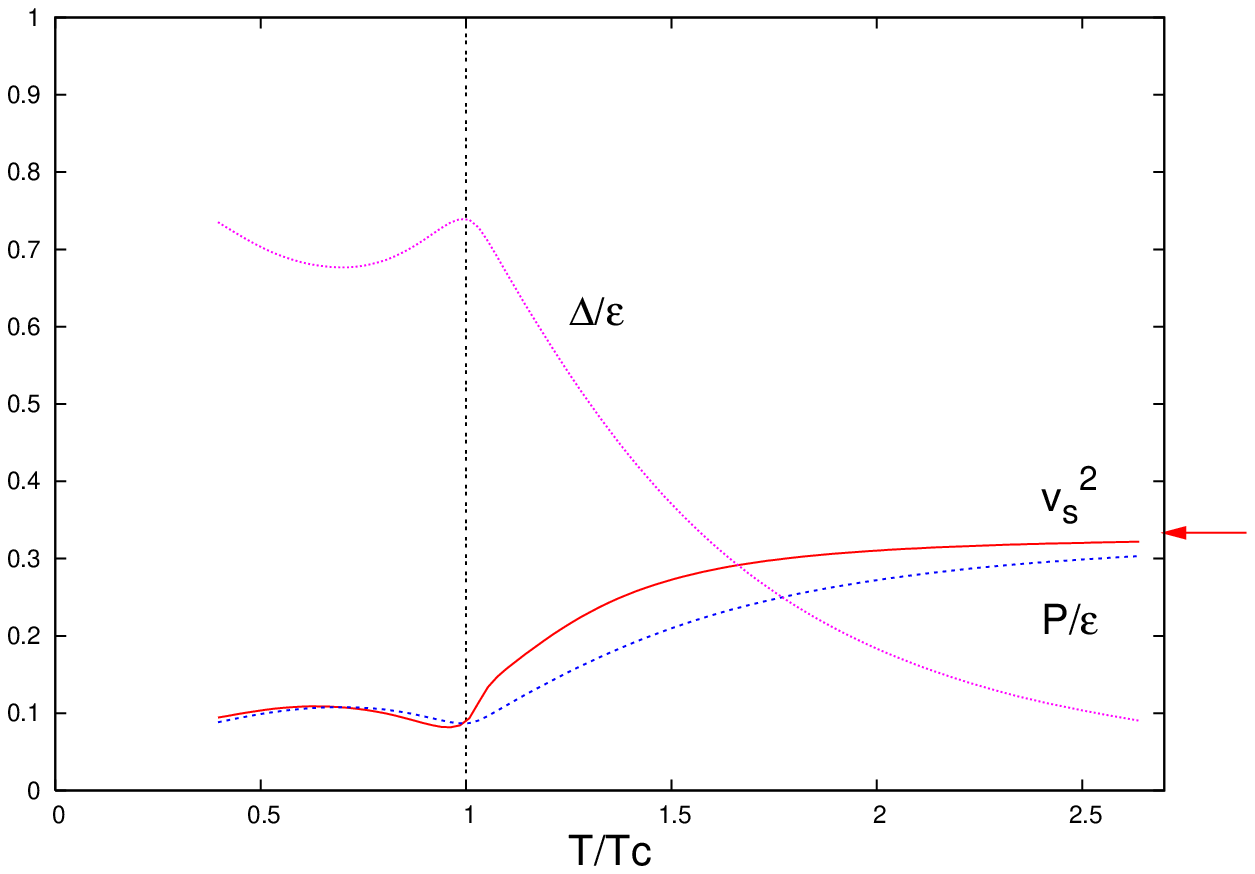}}
\end{minipage}
   \caption{$C_V/T^3$ and $4 \epsilon/T^4$ (left) and Squared velocity 
            of sound $v_s^2$ and conformal measure $\cC=\Delta/\epsilon$ 
           (right) as function of $T/T_c$.
           The arrow on the right shows the ideal gas value.
           }
\label{fg.cv}\end{figure}

Left panel of Fig.\ref{fg.cv}, shows that $C_V$ grows with increasing 
temperature and reaches a peak at $T_c$. 
For comparison, we have also plotted the values of $4 \epsilon/T^4$, 
at which the specific heat is expected to coincide for a conformal gas. 

We plot the speed of sound and conformal measure in right panel of 
Fig.\ref{fg.cv}. The value of $P/\epsilon$ matches with that of $v_s^2$ 
for $T<T_c$ and also goes close again above $2.5 T_c$. But in between 
these two limits the $v_s^2$ is distinctly greater than $P/\epsilon$. 
Thus, $\cC$ would go to zero much faster if we replace $P/\epsilon$ by
$v_s^2$ for computing $\cC$. Near $T_c$ the $v_s^2$ has minimum which 
goes close to 0.08. 

The values for the conformal measure $\cC$ also closely resembles the
Lattice data of Ref.\cite{swa2}. The dip at temperatures less than 
$T_c$ is prominent in both the cases.
At even lower temperatures we find $\cC$ to increase. For a 
nonrelativistic ideal gas, the ratio of $P/\epsilon$ should go to zero,
and thus $\cC$ should then go to 1. On the other hand at high temperatures 
either an ideal gas or a conformal behavior should be recovered for
which $\cC$ should go to zero. 

\section{Possibility of bound states in QGP}
\label{boundstates}
{\it Sanjay K. Ghosh, Munshi G. Mustafa and Rajarshi Ray}
\vspace{0.2in}

Recent LQCD calculations~\cite{Datta,Karsch} 
suggest that pseudoscalars and vectors resonances may exist 
above $T_C$.  
Color bound states  of parton at rest have been claimed by 
analyzing LQCD data~\cite{Shuryak}. The situation, however, changes 
if the test charge, $Q^a$, is in motion relative
to the heat bath. The motion of the particle fixes the direction in space
and spherical symmetry of the problem reduces to axial symmetry. This
implies the loss of spherical symmetry of the Debye screening cloud
around the moving test charge resulting in a wake in the induced charge
due to dynamical screening~\cite{Mustafa05}. The negative minimum found
in the wake potential indicates
a induced space charge density of opposite sign. Thus, a particle moving
relative to a particle in the induced space charge density would constitute
a dipole oriented along the direction of motion. 
We briefly describe the potential due to such dipole interaction in QGP,
details of which can be seen in Ref.~\cite{Mustafa05}.

\begin{figure}
\begin{minipage}[h]{0.48\textwidth}
\centering{\includegraphics[height=1.0\textwidth,width=1.0\textwidth]
{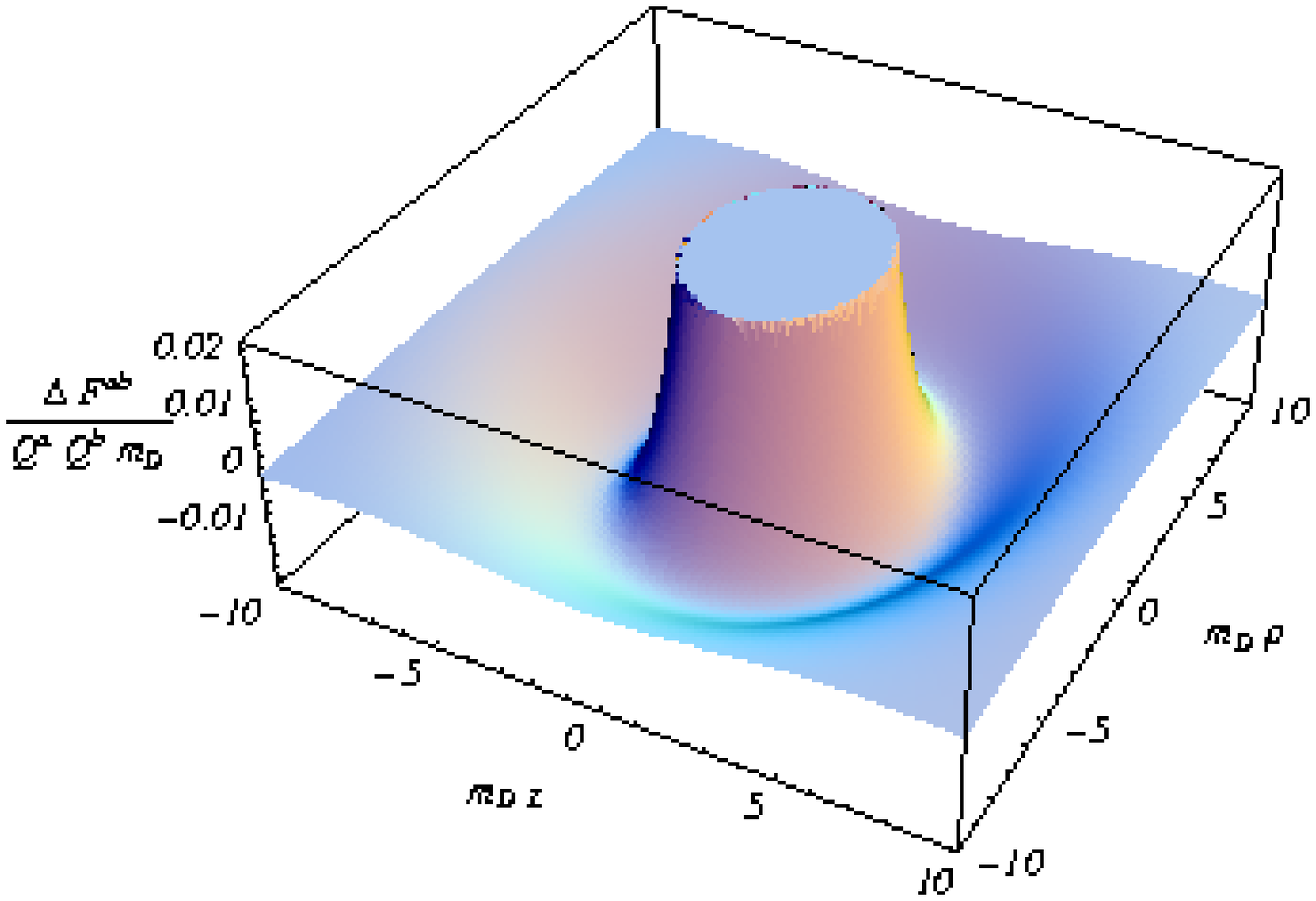}}
\end{minipage}
\begin{minipage}[h]{0.48\textwidth}
\centering{\includegraphics[height=1.0\textwidth,width=1.0\textwidth]
{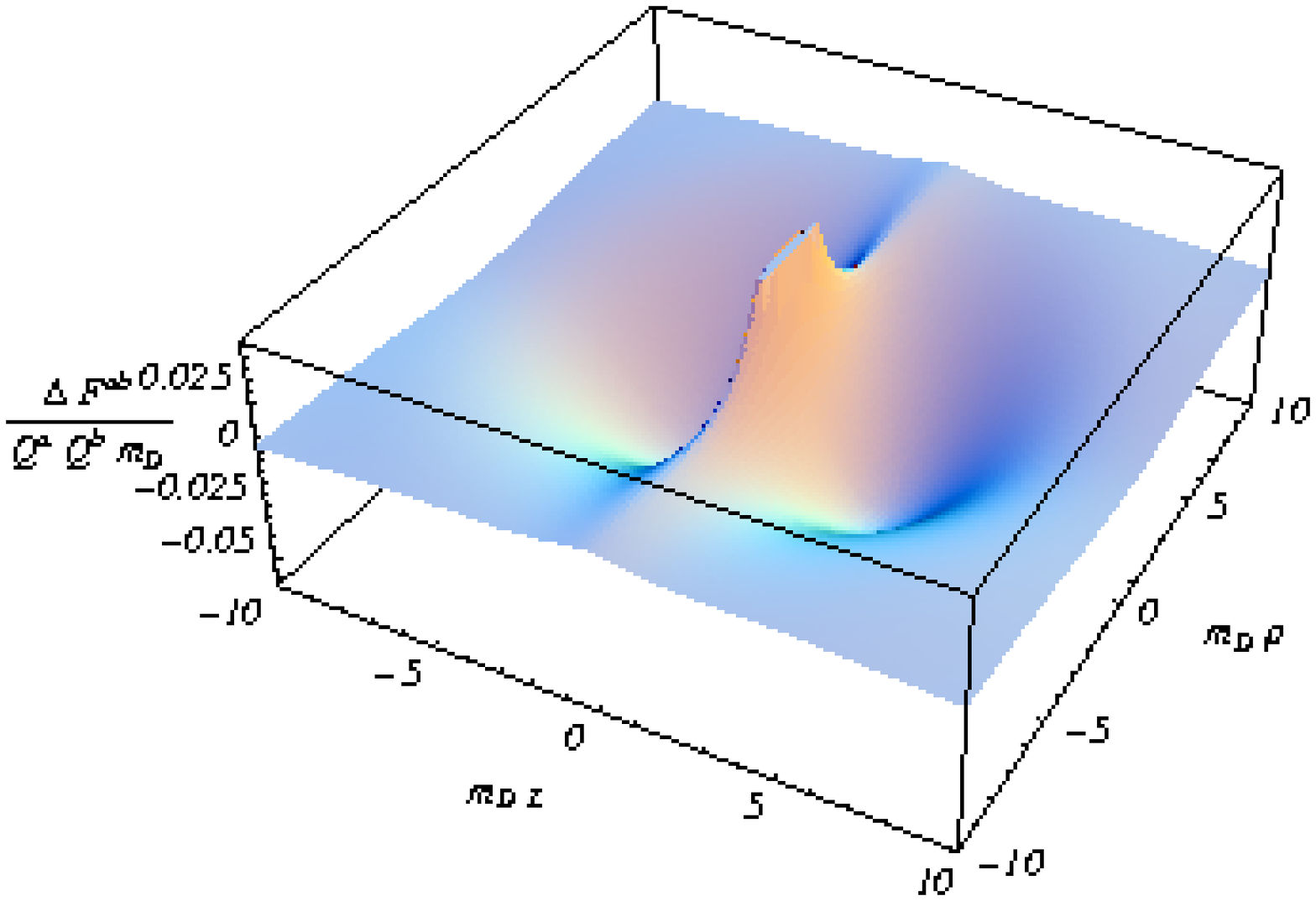}}
\end{minipage}
\caption{Left panel: Spatial distribution of the scaled dipole potential 
with respect Debye screening mass $m_D$ for $v=0.55c$. Right panel:
Same as left panel for $v=0.99c$.  }
\label{fg.dipole}\end{figure}

The spatial distribution of the scaled\footnote{The potential is scaled with
the screening mass $m_D$ as well as with the interaction strength, $Q^aQ^b$. 
So, the details of the potential will depend on the temperature, the strong 
coupling constant and the sign of the interaction strengthi~\cite{Mustafa05}.} 
potential in cylindrical coordinates
are displayed in Fig.~\ref{fg.dipole} for two velocities.
The left panel of Fig.~\ref{fg.dipole} corresponds to $v=0.55c$ where as
the right panel is for $v=0.99c$. 
The dipole potential shows the usual singularity of the screening potential 
at $r=0$ ($z=0$ and $\rho =0$) and a completely symmetric behavior along with a
pronounced negative minimum in the $(\rho -z)$ plane. This is due to the 
compensating effects between the electric and the magnetic interactions. 
For details of electric and magnetic interactions, readers are
referred to \cite{Mustafa05}. 
With the increase of $v$ the relative strength of the dipole potential 
grows strongly.  For $v=0.99c$ a substantial repulsive interaction
has grown in the transverse plane, {\it i.e.}, in the direction of $\rho$
but at $z=0$, becoming responsible for a vertical split of the minimum
in the $(\rho -z)$ plane. However, the general form of the potential mostly 
resembles the Lennard-Jones type with a pronounced repulsive as well 
attractive part.

In QCD the interaction between color charges in various channels is either
attractive or repulsive. A quark and an antiquark yield the sum of irreducible
color representations~\footnote{Their strength of the color
interaction~\cite{Shuryak} can be calculated using $SU(3)$ color group}:
$\mathbf {3 \otimes {\bar 3}= {\bar 1} \oplus 8}$, with the interaction strength of
the color singlet representation is $-16/3$ (attractive) whereas that of
the color octet channel is $2/3$ (repulsive).
Similarly, a two quark state corresponds to the sum of the
irreducible color representations:
$\mathbf {3 \otimes  3 = {\bar 3} \oplus 6}$, where the antisymmetric color
triplet is attractive ($-8/3$) giving rise to possible bound states~\cite{Shuryak}.
The symmetric color sextet channel, on the other hand, is repulsive ($4/3$).
Color bound states ({\it i.e.} diquarks) of partons at rest have been claimed
by analyzing LQCD data~\cite{Shuryak}.  The situation is different
when partons are in motion. The dipole potentials along the
direction of propagation and also normal to it have both attractive and
repulsive parts, similar to the Lennard-Jones form. So, all the
attractive channels or the repulsive channels in the static case get inverted
due to the two comoving partons constituting a dipole in the QGP. This could
lead to dissociation of bound states (or to resonance states) as well
formation of color bound states in the QGP.

Within our model such bound states as well other colored binary states 
in the QGP
can experience different potentials along the dipole direction and the direction
normal to it. Along the direction of motion binary states which were  bound
in QGP may become resonance states or dissociate beyond $T_C$.
Similarly, those colored states which were not bound initially in the QGP,
may transform into bound states.  There are some long distance
correlations among partons in the QGP, which could indicate the
appearance/disappearance of binary states in the QGP beyond $T_C$.
The temperature up to which they survive need further analysis of the bound
states properties in detail, which was initiated during WHEPP-09 and some
progress has been made. The complete work will be reported elsewhere.

\section{$\phi$-production at RHIC}
\label{phi}
{\it Jajati K. Nayak, Jan-e Alam, Bedangadas Mohanty, Pradip Roy, Abhee K. Dutt-Mazumder}

\vspace{0.2in}
 Among the promising signals to analyse the matter formed in Relativistic
Heavy Ion Collision, the study of strangeness is important one. The 
production and evolution of strange particles give good information about
the formation of QGP. $\phi$ meson is a bound state of strange 
quark(s) and anti strange quark ($\bar{s}$). The interaction of $\phi$
meson with nuclear matter is suppressed according to Okubo-Zweig-
Izuka(OZI) rule. $\phi$ meson after its production during hadronisation
suffers less rescattering with hadronic matter. So it gives thermodynamic
information of the state of matter during hadronisation stage. From the 
$\phi$ spectra at $\sqrt{s}$=200 GeV (RHIC energy) we extract the information
about critical temperature $T_c$ of the quark-hadron phase transition.

\begin{figure}[htbp]
\epsfxsize=6cm
\centerline{\epsfbox{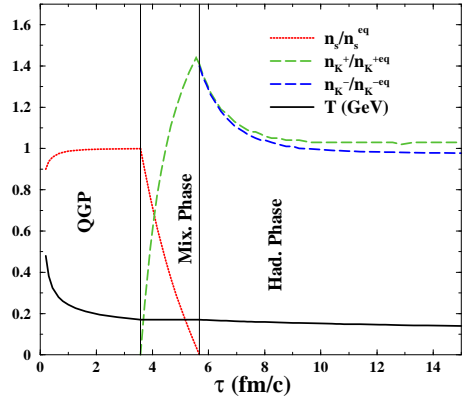}}
\caption{Evolution of strange particles and temperature when
$T_c$=170 MeV.  $n_i/n_i^{eq}$ is the ratio of no. density of i-type
particle. }
\label{figure1}
\end{figure} 

 We assume a first order quark-hadron phase transition for the RHIC 
energy $\sqrt{s}$=200 GeV. In QGP phase s and $\bar{s}$ quarks are 
produced mainly by gluon gluon fusion and annihilation of light
quarks and anti-quarks. These s and $\bar{s}$ forms $\phi$ through
hadronisation process. This production is not OZI suppressed.
We expect excess $\phi$ mesons if QGP is formed in the 
initial state of Heavy Ion Collisions as compared to a hadronic initial 
state. The production of $\phi$ from sources other than plasma is very small.
$\phi$ spectra is experimentally reconstructed from Kaons since 
it decays to $K^{+}$ and $K^{-}$. Although $K^{+}$ and $K^{-}$ 
carry some scattering effect still $\phi$ meson spectra gives 
reliable information of the thermodynamic state of matter during 
its formation.

We calculate the ratio of the multiplicity of $\phi$ meson to 
the total multiplicity at mid rapidity of RHIC experiment 
from thermal model. The ratio depends on the critical temperature 
$T_{c}$ and the effective degeneracy $g_{eff}$. We took these value 
from lattice QCD calculation. We found the measured value 
to be more than the calculated one. This is expressed as a over 
production or enhancement factor $\gamma_{\phi}$. To know where 
and when this over production occurs we adopted Boltzmann equation 
and solved for the evolution of strange particles(e.g, s,$\bar{s}$,
$K^{+}$,$K^{-}$ etc)  in QGP, mixed and hadron phases. With proper 
initial conditions from hadron spectra we solved the Boltzmann 
equation and temperature evolution equation and found the over 
production factor $\gamma_{\phi}$ $~$ $r_{K^+}^2$=1.63 ($r_{K^+}$ 
being the ratio of non-equilibrium to equilibrium density of $K^{+}$) when 
$T_{c}$=170 MeV. The production of $K^+$ is restricted by a parameter
$\delta$. Maximum half of the strange quarks can go to the $K^+$. 
This is expressed as $\delta$=0.5. So this puts the limit on 
the critical temperature i.e, $T_c$ $>$ 170 MeV. Similarly a limit on 
$g_{eff}$ also has been put~\cite{jajati}.

\section{AdS/CFT $\longrightarrow$ QCD/QGP ?}
\label{Ads}
{\it Balram Rai}
\vspace{0.2in}

\noindent The idea that QGP at RHIC, $T\sim 2T_C$, seems to be in a 
strongly coupled
regime (interaction energy $>>$ kinetic energy) and has presently been
attracted a lot of attention :  
\begin{enumerate}
\item The equation of states obtained in LQCD deviates from ideal gas 
behavior. Combining the LQCD data on quasiparticle masses and interparticle
potential one indeed finds~\cite{Shuryak} a lot of bound states and 
resonances above $T_C$.  This could be a explanation for charmonium states 
remaining bound up to (2-3)$T_C$, as was directly observed on 
lattice~\cite{Datta}. However, resonances above $T_C$ due to marginal bound 
states give rise to large cross-section and small mean free path leading to 
collective flow.
\item Collective phenomena observed at RHIC lead to view QGP as a near perfect
fluid~\cite{fluid}. The deviation of elliptic flow data from the prediction 
of hydrodynamic calculations only occur at $p_T \sim 1.5-2$ GeV/$c$ and this 
leads to viscosity to entropy ratio, $\eta/s\sim$ 0.1-0.2, 
which is more than order of magnitude less than that in pQCD.
\item The charm diffusion constant~\cite{Moore} deduced from the single 
electron data and elliptic flow is also an order of magnitude less than 
pQCD estimates. 
\item The interaction parameter $\Gamma= \langle {\rm{ potential\ \ energy}}
\rangle/T$ is obviously not small for QGP~\cite{Thoma}. At such $\Gamma$
the classical strongly coupled plasma is a good liquid. 
\end{enumerate} 
As seen the pQCD calculations of transport coefficients are not
reliable around and above $T_C$, which is believed to be in strong coupling 
regime. On the other hand the LQCD is a rigorous calculational method 
applicable in hot, strongly coupled, gauge theory. Because it is formulated
in Euclidean space, it is well suited  for computing static thermodynamic
quantities and less well-suited for transport coefficients, or dynamical
processes of any sort. Complementary nonperturbative techniques are thus 
desirable. One such technique is the Anti-de Sitter/Conformal Field Theory
(AdS/CFT) correspondence, which maps nonperturbative problems in certain
hot strongly coupled gauge theories onto calculable problems in a dual 
gravity theory~\cite{Ads}. This method has been applied to calculate
the shear viscosity~\cite{viscosity} in ${\cal N}=4$ supersymmetric 
Yang-Mills (SYM) theory which gives an upper bound of $\eta/s\sim 1/4\pi$,
close to the RHIC value. Also effort~\cite{diffusion} was made to compute
certain diffusion coefficients in the same spirit. The ratio of
pressure to that of Stefan-Boltzmann limit in ${\cal N}=4$ SYM is
remarkably close to the corresponding ratio in LQCD at temperatures 
a few times $T_C$ where it is strongly coupled~\cite{pressure}. 

Usually the properties of ${\cal N}=4$ SYM theory are completely
different from QCD. The former one is a conformal theory with no particle
spectrum or ${\cal S}$-matrix, whereas the latter one is confining theory
with a realistic particle interpretation.  In view of 
this members of this working group were interested to know the basics of 
AdS/CFT correspondence to supersymmetric gauge theories in strong coupling 
limit and their connection to QCD/QGP.  Balram Rai delivered two comprehensive 
lectures on this topic in a very elementary level, which are discussed below 
in brief.

\begin{enumerate}
\item[$\bullet$] 
${\cal N}=4$ SYM is a conformally invariant theory with two parameters:
the rank of the gauge group $N_C$ and the t'Hooft coupling 
$\lambda=g^2_{\rm{YM}}N_C$ with $g_{\rm{YM}}$ is Yang-Mills coupling.
Its on-shell field content includes eight bosonic and eight fermionic degrees
of freedom, all in color adjoint representation.

\item[$\bullet$] 
Basic argument leading to this correspondence: consider in type IIB
string theory, a configuration of $N_C$ $D3$-branes stacked on top of each 
other.
The appropriate theory living on the branes is ${\cal N}=4$ SYM theory. On
the other hand if $N_C$ is large, the stack of $D3$-branes has large 
tension, which curves space-time. In the limit of large t'Hooft coupling 
$\lambda$, the brane geometry has small curvature  and can be described
by supergravity. Therefore, one can have two descriptions of the same physics
in terms of strongly coupled gauge theory on the branes and classical 
gravity on a certain background. 

\item[$\bullet$] 
AdS/CFT conjecture (or gauge/string duality) states
that this theory is exactly equivalent to type IIB string theory in
an $AdS_5\times S^5$ gravitational background, where $AdS_5$ is
a five dimensional anti-de Sitter space and $S^5$ is five dimensional
sphere. At large $N_C$ and large $\lambda$, the string theory can be
approximated by classical type IIB supergravity. 

\item[$\bullet$] 
The above approximation permits nonperturbative calculations in quantum 
field theory
mapped into problems in classical general relativity. In this context, 
raising the temperature of the gauge theory corresponds to a black hole (or
a black brane) into the center of $AdS_5$. According to AdS/CFT, the 
Hawking temperature of the black hole becomes the temperature of the gauge
theory.
\item[$\bullet$] 
Even at $T=0$, the  ${\cal N}=4$ SYM theory  is a conformal one whereas the 
QCD is confining theory, but $T\neq 0$ both theories describe hot, nonabelian
plasmas with Debye screening, finite spatial correlation length, and
qualitatively similar hydrodynamic behavior. The major differences is that
all the excitations (quarks, gluons and scalars) in ${\cal N}=4$ SYM plasma
are in adjoint representation, while hot QCD plasma has quarks in fundamental
and gluons in adjoint representation. 
This leads to an assumption  that the fundamental
representation fields have negligible influence on bulk properties of the 
plasma. One may view the quarks as test particles which serve as probes of 
dynamical processes in the background of ${\cal N}=4$ plasma. 

\item[$\bullet$] As found the shear viscosity, the thermodynamic quantities
such as pressure, energy density, entropy etc. computed
in ${\cal N}$ SYM in strong coupling limits (large $N_C$ and large $\lambda$)
are insensitive to details of the plasma composition or 
the precise interaction strength. These values are remarkably close to those
properties of QGP observed in RHIC and obtained LQCD.

\item[$\bullet$] 
These observations have led to think a connection of AdS/CFT gravity dual to
RHIC collisions !

\end{enumerate}

\section{Neutrino emission rates in crystalline 
color superconducting quark matter} 
\label{crystal}

{\it Prashanth Jaikumar, Hiranmaya Mishra and Andreas Schmitt}


\vspace{0.2in}

The existence of deconfined quark matter inside
 compact stars can be verified if the stellar cooling rate is found to
 be substantially different than for neutron stars, and can be
 attributed directly to the neutrino emission rates of
 (superconducting) quark matter. This requires\footnote{This work was initiated
during WHEPP-09 and substantial progress has already been made.}
\noindent 1) identifying the ground state of dense QCD at intermediately large
quark chemical potentials ($\mu_q\sim 300-400$ MeV) that characterize the
interior of neutron stars, and
\noindent 2) computing the neutrino emission rate and specific heat of this phase to determine its cooling behavior.

\noindent One possible manifestation of the diquark phase at
 intermediate densities, where the strange quark mass is large, is a
 crystalline color superconductor in which quarks with different Fermi
 surfaces pair at non-zero momentum, resulting in an inhomogeneous but
 spatially periodic order parameter~\cite{KR1}. This phase
 spontaneously breaks translation and rotational symmetries, and the
 free energy of the system is minimized when the gap varies spatially
 in accordance with the residual discrete symmetries of this phase.
 This is the QCD incarnation of the LOFF
 (Larkin-Ovchinnikov-Fulde-Ferrell) phase in electronic spin
 systems. 


\noindent The presence of color superconducting phases is expected to alter the 
cooling behavior of neutron stars. 
The equation that  determines the cooling of an isothermal neutron star of volume $V$ in global heat balance is

\beq
c_v\frac{dT}{dt}=-L_{\nu}\equiv -\epsilon_{\nu}V
\eeq
\noindent where $T$ is temperature, $t$ is time, $\epsilon$ is the neutrino
emissivity and $c_v$ the specific heat at constant volume $V$ of the quark
phase.
Minimization of the thermodynamic potential leads to the nonisotropic
gap equation~\cite{KR2}

\beq
\Delta=\frac{2G}{(2\pi)^3}\int d p\frac{2\sin^2\frac{\beta}{2}}{
\sqrt{\left(| q+ p|+| q- p|-\bar\mu\right)^2+4\Delta^2
\sin^2\frac{\beta}{2}}}
\label{gapeq}
\eeq

where the momentum integration runs over the pairing region defined by the domain \{${\cal P}:{\bf p}| E_1(p)>0, E_2(p)>0$\}. Here, the angle $\beta$ is the angle between the up quark momentum
($q+p$) and the down quark momentum ($ q- p$).
Further, the quasi particle energies are given as~\cite{JB}
\bea
E_1(p)&=&+\delta\mu +\frac{1}{2}(| p+ q|-|p- q|) \nonumber \\
&& +\frac{1}{2}\sqrt{\left(| q+ p|+| q- p|
-2\bar\mu\right)^2+4\Delta^2\sin^2(\frac{1}{2}\beta)}
\label{e1}
\eea
and 
\bea
E_2( p)&=&-\delta\mu -\frac{1}{2}(| p+ q|-| p- q|)\nonumber \\
&& +\frac{1}{2}\sqrt{\left(| q+ p|+| q- p|
-2\bar\mu\right)^2+4\Delta^2\sin^2(\frac{1}{2}\beta)}
\label{e2}
\eea
In the above, $\bar\mu=(\mu_u+\mu_d)/2$ and $\delta\mu=\mu_d-\mu_u$ is
the average and the difference respectively of the chemical potentials
of the two condensing quarks.  Thus, crystalline superconductivity
is characterized by dispersion relations which vary with direction
of momentum, yielding gaps which vary from zero up to a maximum of
$\Delta$.

\begin{figure}[htpb]
\begin{minipage}[h]{0.4\textwidth}
\hspace*{2.7cm}{\includegraphics[height=1.2\textwidth,width=1.6\textwidth]
{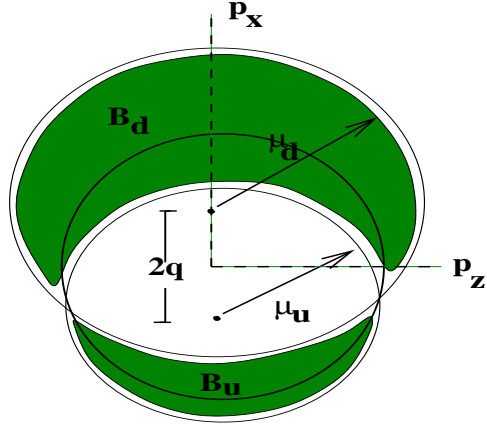}}
\end{minipage}
\caption{Allowed (unshaded) and blocked (shaded) regions for pairing between up and down quarks in the LOFF phase. The 2-dimensional projection of the Fermi spheres for the two flavors of quark are shown displaced by an amount $2{\bf q}$. The dispersion $E_1(E_2)$ vanishes on the boundary of the lower (upper) banana. These surfaces make the principal contributions to the specific heat and neutrino emissivity. Particle-hole pairing occurs on the boundary of the elliptical region but is clearly disfavored at low temperatures. \label{loff}}
\end{figure}

For the purpose of neutrino emissivity calculations, it is the nodal
surfaces described by $E_1=0$ and $E_2=0$ that make the dominant
contributions to the phase space integral in Eqn.(\ref{nuemiss}).
Thus one needs to know the dispersion relation near the 'blocking'
regions $B_u,B_d$ in Fig. \ref{loff}, whose boundary can be specified by solving
$E_i(p)=0$ for the angle between the condensate momentum $q$ and
the relative momentum between the pairing quarks $ p$.

The neutrino emissivity is calculated from

\bea
\label{nuemiss}
\epsilon_{\nu}&=&\left[\prod_{i=1}^{4}\int_{E(p_i)=0}\frac{d^3p_i}{(2\pi)^3}
\right]E_{\nu}Wn(p_d)(1-n(p_u))(1-n(p_e))\nonumber \\
&&\times \psi_d(p_d)^2B\psi_u(p_u)^2
\eea

where $i=1..4$ represents $d,u,e,\nu$ respectively.

\beq
W=\frac{(2\pi)^2\delta^4(p_d-p_u-p_e-p_{\nu})}{\prod_{i=1}^{4}2E_i}|M|^2
\eeq

with $|M|^2=64G_F^2{\rm cos}^2\theta_c(p_d.p_{\nu})(p_u.p_e)$ being
the matrix element for the weak interaction process. The $\psi$'s are
Bogoliubov coefficients for the ($du$) quasiparticles.

The specific heat can be calculated from

\beq
c_v=\frac{1}{T^2}\int\frac{d^3p}{(2\pi)^3}\frac{E(p)^2}
{(e^{E(p)/T}+1)(e^{-E(p)/T}+1)}
\eeq

Since we are interested in the low-temperature behavior of the specific heat, we can approximate

\beq
c_v\approx\frac{1}{T^2}\int\frac{d^3p}{(2\pi)^3}E(p)^2e^{-E(p)/T}
\eeq


The temperature dependence of the specific heat can be classified with respect to the behavior of the dispersion relations in the vicinity of its zeros and the dimension $d=0$ (point), 1 (line) and 2 (surface) of the sub-manifold where the dispersion vanishes. Using the above formula, we can deduce the results for the temperature dependence of the specific heat in the following table:

{\centering \begin{tabular}{|c|c|c|c|}
\hline
$c_v$ & $0$ & $1$ & $2$
\\ \hline \hline
linear & $T^3$ & $\bar{\mu}T^2$ & $\bar{\mu}^2T$ \\ \hline
quadratic & $\bar{\mu}T^2$ & $\bar{\mu}^{3/2}T^{3/2}$ & $\bar{\mu}^{5/2}T^{1/2}$\\ \hline
\end{tabular}\par}


where constants of proportionality of order one have been omitted. These relations indicate that the circle of nodes at the tips of the two banana-shaped regions contribute the most to the specific heat since $\bar{\mu}\gg T$. The point nodes contribute less to the specific heat but will dominate the contribution to the neutrino emissivity since their phase space weights are larger. The computation for the emissivity is complicated by the need to perform a bounded phase space integration excluding the blocking regions, and this will be attempted numerically. 
\section*{Acknowledgments}

We are grateful for the hospitality provided by the Institute of Physics, Bhubaneswar 
and the organizers of WHEPP-09  for providing the stimulating environment for 
working group discussions.
 

\end{document}